\documentclass[aps,onecolumn,floatfix]{revtex4-1}
\usepackage{bbold}
\usepackage{graphicx}
\usepackage{amsmath}

\usepackage{color}
\usepackage{lineno}

\newcommand{\matA}{\underline{\mathbf{A}}}
\newcommand{\matB}{\underline{\mathbf{B}}}
\newcommand{\matC}{\underline{\boldsymbol{\mathcal{C}}}}

\begin{document}
\title[]{Rayleigh-Bloch, topological edge and interface waves\\ for structured elastic plates}
\author{G.~J. Chaplain$^1$, M.~P. Makwana$^{1,2}$, and R. V. Craster $^1$}
\affiliation{$^1$ Department of Mathematics, Imperial College London 180 Queen's Gate,\\ South Kensington, London SW7 2AZ}
\affiliation{
 $^2$ Multiwave Technologies AG, 3 Chemin du Pr\^e Fleuri, 1228, Geneva, Switzerland}
 
\begin{abstract}
Galvanised by the emergent fields of metamaterials and topological wave physics, there is currently much interest in controlling wave propagation along structured arrays, and interfacial waves between geometrically different crystal arrangements. We model array and interface waves for structured thin elastic plates, so-called platonic crystals, that share many analogies with their electromagnetic and acoustic counterparts, photonic and phononic crystals, and much of what we present carries across to those systems. These crystals support several forms of edge or array-guided modes, that decay perpendicular to their direction of propagation.   To rapidly, and accurately, characterise these modes and their decay we develop a spectral Galerkin method, using a Fourier--Hermite basis, to provide highly accurate dispersion diagrams and mode-shapes, that are confirmed with full scattering simulations. We illustrate this approach using Rayleigh Bloch modes, and generalise high frequency homogenisation, along a line array, to extract the envelope wavelength along the array. Rayleigh-Bloch modes along graded arrays of rings of point masses are investigated and novel forms of the rainbow trapping effect and wave hybridisation are demonstrated. Finally, the method is used to investigate the dispersion curves and mode-shapes of interfacial waves created by geometrical differences in adjoining media.
\end{abstract}
\maketitle
\section{Introduction}

\label{Intro} 
During the past two decades, there has been much interest in the field of metamaterials \cite{maier17a} and waves through structured media in the context of photonic crystals \cite{joannopoulos08a}. 
Metamaterials are artificial materials, originating in optics and photonics, that exhibit exotic properties, often not occurring in nature such as negative refractive index \cite{veselago68a}, allowing the novel control of wave propagation \cite{smith00a,pendry00a}. More recently, it has been recognised that metamaterial ideas based around local resonance to create sub-wavelength devices are ubiquitous across wave physics and they have been translated from electromagnetic systems into acoustic, seismic and elastic regimes \cite{craster2012acoustic,deymier13a,colombi17a,craster17a}.

The closely related area of photonics, in optics, and their acoustic and elastic analogues of phononic crystals also uses structured media to control waves although now through Bragg interference. Both photonics and metamaterials use structuration, often periodic, of the medium, or of a surface, to control waves and it is desirable, for instance, to channel wave energy to precise positions with minimal energy loss; this can be achieved using waveguiding within crystals by removing elements \cite{mekis96a}, or by making geometrical changes in adjacent media to create guiding along their interface   \cite{lu_dirac_2014-1,khanikaev_two-dimensional_2017}. An alternative way to guide energy along surfaces is to use line, or surface, arrays and there exist array guided waves that go under various different names: spoof surface plasmons in physics \cite{pendry04a,fernandez11a}, Yagi-Uda antennas and related electromagnetic guiding \cite{hurd54a,sengupta59a}, Rayleigh-Bloch (RB) waves \cite{porter99a} in acoustics, and edge waves \cite{evans93a} in water waves. We call these array guided waves RB waves henceforth and they owe their existence to the periodicity of the array. Our interest here is in the efficient modelling of these various guided wave systems and of particular interest is their model shape and frequency dependence via dispersion curves. These array guided waves, interface states, and edge modes, share much commonality in that one can formulate the underlying problem in an infinite strip and then search for eigenstates that decay exponentially away from the propagation direction. 

\begin{figure}[!hb]
    \centering
    \includegraphics[scale = 0.25]{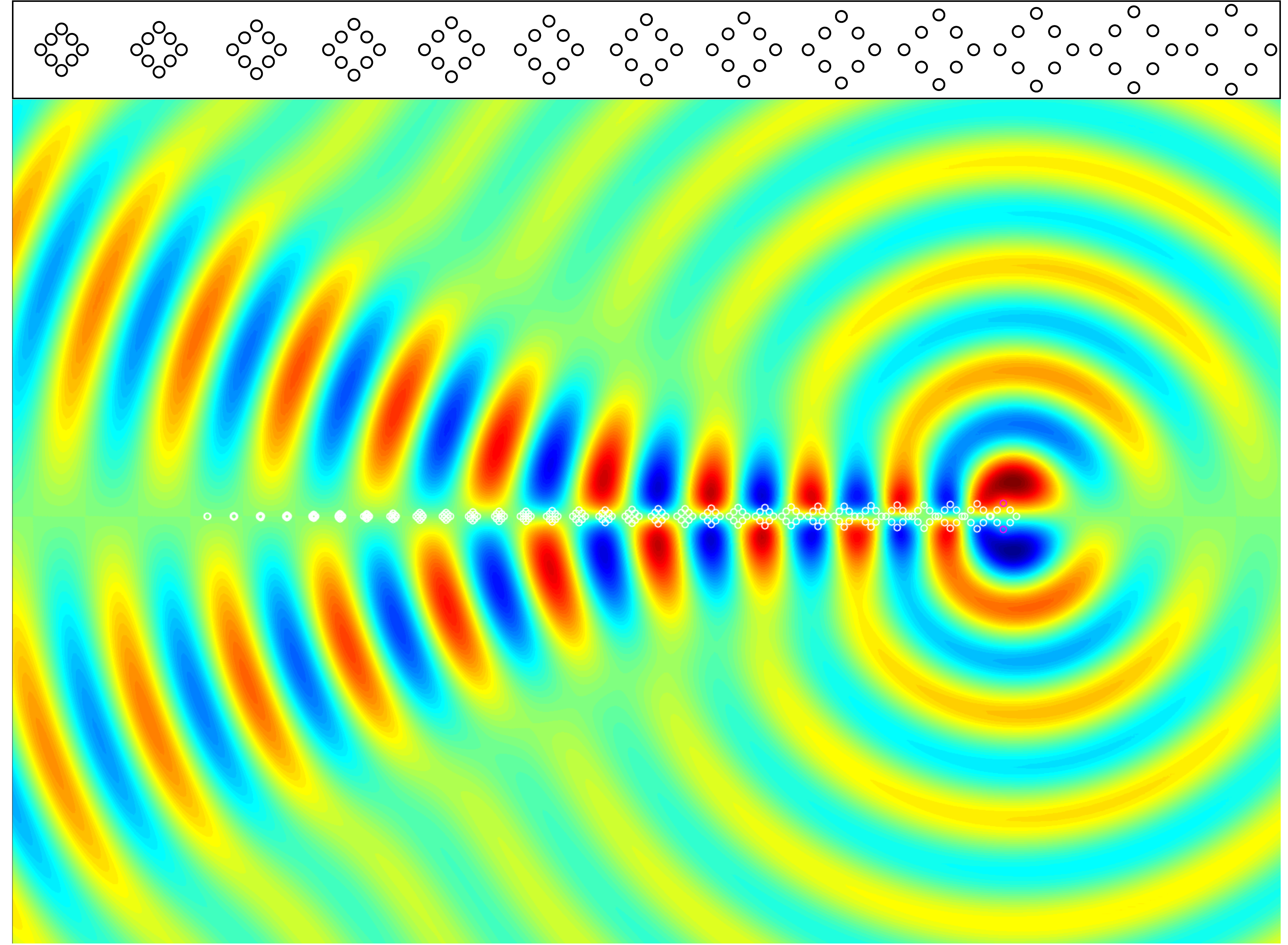}
    \caption{The hybridisation of a RB wave along a graded array converting into a bulk flexural wave as excited with a point dipole source. An expanded view of a section of the grading profile, as shrinking diamond arrangements of mass-loaded points, is shown in the upper inset. In this simulation the mass value is  $10$, and excitation frequency is $0.7972$ with array pitch $2$.}
    \label{fig:Fig1}
\end{figure}

Once the guided waves have been accurately identified then there are a range of interesting physical effects, for instance those associated with grading arrays, or crystal interfaces, to enable energy harvesting \cite{bennetts18a}, that can be explored; we do so here in the context of rainbow trapping and mode hybridisation. In particular we exploit the symmetries of the guided waves to create an array that switches between trapping and hybridisation.  
For these graded arrays one adiabatically alters the array to engineer cut-offs and band gaps to create spatial frequency selection and trap the wave; this effect has been realised in optics for some time \cite{Tsakmakidis2007,Gan} and subsequently in acoustics \cite{Romero-Garcia2013,Zhu2013} and seismology \cite{Colombi2016}; for seismic elastic waves  mode conversion between surface and bulk waves occurs which is of interest for controlling ground vibration. Our aim here is to demonstrate that one can also mode convert in the elastic plate setting by appropriately choosing the grading and geometry of the defect within the strip, and taking advantage of the symmetries of the modes; this being an exemplar of the design possibilities given via the methodology we employ. An example of the hybridisation of a RB wave into a propagating flexural wave is shown in Figure~\ref{fig:Fig1}. The prediction of this wave beaming, and features associated with it,  follows from the coupling capabilities between odd and even mode shapes, which are easily obtained from our modelling of the perfectly periodic system. 

We concentrate here upon these guided waves in the setting of flexural waves propagating upon thin elastic plates, using the Kirchhoff-Love equation (KL) \cite{graff75a}, which have been much studied  \cite{Farhat2010, Haslinger2012, Smith2012, Torrent2013,climente16a}, both in terms of their practical importance and in terms of being a convenient model to explore geometry \cite{Pal2017,Makwana2018a} without requiring heavy numerical simulations, as the model benefits from elegant analytical solutions for point loading \cite{evans2007penetration}; the Green’s function is, attractively, non-singular and remains bounded. Moreover, thin elastic plates have been identified theoretically as candidates which can support topological edge or interface states \cite{Torrent2013,Pal2017,Chaunsali} as well as experimental observation \cite{Vila2017}. It is therefore important to be able to characterise the decay of these guided edge states in these exotic new materials to assess their feasibility for engineering applications.

Conventionally, to extract the decaying eigensolutions, a ribbon of material is taken sufficiently long such that the exponentially decaying solutions are not impacted by the boundary conditions at the top, or bottom, of the ribbon, and Floquet-Bloch conditions are applied to the sides of the ribbon; one can then use many numerical methods (finite element, finite difference or spectral collocation) to solve the ensuing eigen-problem in the physical domain. However, this has the disadvantage that many spurious modes are detected as the ribbon is infinite, and not finite, and these spurious modes are valid solutions of the truncated numerical problem but not of the physical one. It can also become time consuming to extract weakly decaying modes since the truncated finite  strip needs to be very long. We often want rapid solutions, only for the decaying modes that we care about, particularly when designing and optimising geometries and material parameters and so a scheme designed to extract only the decaying solutions is attractive. 

For doubly-periodic media, such as a conventional photonic crystal \cite{joannopoulos08a}, the plane--wave--expansion (PWE) method \cite{johnson01a} is a spectral method that is particularly attractive and has been used in a variety of related fields with much success including for KL plates \cite{Chaunsali,xiao2012flexural}. It also has the advantage that the numerical work is performed in the reciprocal domain, i.e. directly in Fourier space. One can, of course, treat a truncated long finite ribbon just as a tall cell in a doubly-periodic medium and use this approach to generate the eigensolutions for the edge states as in \cite{Makwana2018a}. 
However, to calculate the dispersion diagrams for decaying solutions, such as Rayleigh--Bloch (RB) waves in \cite{evans2007penetration,porter2005embedded}, For the topological edge states  \cite{Mousavi2015,Guo2017,Li2018}, it is ultimately desirable to have a method that specifically finds only decaying solutions, and quickly characterises the dispersion curves.

We choose to adopt a Fourier--Hermite basis expansion of the wave field in a unit cell of the form of an infinite strip; the Hermite expansion builds in the desired decay and we describe this methodology in Section~\ref{FHG}, where the subtleties of the expansion are explained, as well as the development of the numerical model. In Section~\ref{RB}, numerical results are shown for a variety of line arrays displaying applications of our approach to rainbow trapping and hybridisation effects. Furthermore, the interface states found in adjoining hexagonal structured media are discussed, as is the numerical implementation of high frequency homogenisation. Finally, in Section~\ref{Conc} conclusions are drawn and motivation for further research presented.

\section{Fourier--Hermite Galerkin Spectral method} \label{FHG}
\label{Galerkin}
The essence of the Galerkin spectral method, applied to flexural waves on elastic plates, is to transform the physical problem modelled by the KL governing equation \eqref{eq:KL} into matrix algebra in reciprocal space.   This is achieved by expanding the wave field with a suitable basis that encapsulates the underlying behaviour of the system, that is quasi-periodic Floquet-Bloch conditions in one direction, along the array, and decay in the other. For an infinite, thin elastic plate in the Cartesian \((x,y)\) plane,  loaded at points $\mathbf{x}_{N}^{(j)}$ that produce a reaction force, the governing KL equation (assuming time-harmonic behaviour) takes the form \cite{graff75a}

\begin{equation}
\nabla^{4}w(\mathbf{x}) - \Omega^{2}w(\mathbf{x}) = \sum\limits_{N}\sum\limits_{j = 1}^{N_{e}} F_{N}^{(j)}\delta(\mathbf{x}-\mathbf{x}_{N}^{(j)}),
\label{eq:KL}
\end{equation}
where \(w(\mathbf{x})\) is the plate displacement at \(\mathbf{x} = (x,y)\). We assume geometric periodicity in \( x \) which leads to a mathematical problem given in terms of an infinite strip, where \(N\) indicates summing over each strip. The remaining notation in \eqref{eq:KL} is that \(N_{e}\) gives total number of the point loadings within each strip and \(x_{N}^{(j)}\) gives the position of the \(j^{th}\) element within the \(N^{th}\) strip, with \( F_{N}^{(j)} \) being the forcing by such an element. The elements can take the form of clamped pins, masses or resonators, which will ultimately influence the form of \(F\)  \cite{evans2007penetration}, and we fix upon using mass loading throughout this article. A time-harmonic response is implicitly assumed, such that the plate parameters are absorbed into a non-dimensionalised frequency, \(\Omega\), with \(\Omega^2 = \rho h\omega^2/D\), where \(\rho\) is the mass density of the plate and \(h\) is the plate thickness, with \(\omega\) being the dimensional angular frequency. \(D\) is the flexural rigidity, which encodes the Young's modulus, \(E\), and Poisson's ratio, \(\nu\), of the plate through \(D = Eh^3/12(1-\nu^2)\). For simple mass loading the point forcing, \(F\) in \eqref{eq:KL}, is proportional to the displacement of the plate at that point, \(w_{N}^{(j)}\). 

We take advantage of the geometric periodicity in the physical problem, we take each infinite strip to have width of $a$ and we can fix attention upon a single strip in $\vert x\vert \leq a/2$ with a phase shift $\kappa$, the Bloch wavenumber, across the strip such that $w(a/2,y)=\exp(ai \kappa) w(-a/2,y)$;  the dispersion relations that we seek relate the frequency to this Bloch wavenumber.  
 Incorporating these Bloch conditions, Eq. \eqref{eq:KL} becomes 
\begin{equation}
\nabla^{4}w(\mathbf{x}) - \Omega^{2}w(\mathbf{x}) = \Omega^2\sum\limits_{N,j} \exp[{-i\kappa(x_N^{(j)} - x_{\mathbb{1}}^{(j)})}]M^{(j)}w_{\mathbb{1}}^{(j)}\delta(\mathbf{x}-\mathbf{x}_{N}^{(j)}).
\label{eq:KLBloch}
\end{equation} 
where \(M^{(j)}\) is the mass value of the \(j^{th}\) point mass in the strip, and we introduce the fundamental strip, denoted by $\mathbb{1}$. 

Motivated by the behaviour of the solutions that we aim to capture within the infinite strip, we expand the wavefield in a Fourier--Hermite basis, such that propagating behaviour is captured along the array by the Fourier expansion and the exponential decay is captured by the Hermite expansion. 
 
This choice of basis functions results in the wavefield taking the form
\begin{equation}
w(\mathbf{x}) = \sum\limits_{n=-\infty}^{\infty}\sum\limits_{m=0}^{\infty}W_{nm}\exp[{i(G_{n}-\kappa)x}]\psi_{m}\left((y-\Delta)/\tau \right).
\end{equation}
The unknown coefficients, \(W_{nm}\), are  determined in the Galerkin scheme and are used to  give the modeshapes.  The coefficient \(G_{n}\) is the element of the reciprocal lattice vector in the \({{x}}\) direction such that \(G_{n} = 
{2\pi n}/{a}, 
 \)
 which defines the one-dimensional irreducible Brillouin zone, where  \({a}\) is the width of the fundamental strip. The major change from a standard plane wave expansion is the introduction of \(\psi_{m}((y-\Delta)/\tau)\) which is the (symmetrically weighted) orthonormal Hermite function, \cite{boyd2017dynamics}, of order \(m\) as given below in \eqref{eq:orthonormal}.  This Hermite representation introduces a scaling factor, \(\tau\), and shifting parameter, \(\Delta\), that, as we shall see later, can be of importance for fast convergence of the expansion. This Galerkin scheme leads to a generalised eigenvalue problem, that finds the frequencies given the Bloch wavenumber, as the eigenvalues and with the $W_{nm}$ coefficients as the corresponding eigenvector, where the infinite sums are truncated.

\subsection{Development of the Method}\hspace*{\fill} \\
The scaled orthonormal Hermite functions of order \(m\), \(\psi_{m}((y-\Delta)/\tau)\), are related to the Hermite polynomials, \(H_{m}((y-\Delta)/\tau)\) \cite{abramowitz1965handbook}, through 
\begin{equation}
\psi_{m}((y-\Delta)/\tau) \equiv (2^mm!\tau\sqrt{\pi})^{-1/2}\exp[{-(y-\Delta)^2/2\tau^2}]H_{m}((y-\Delta)/\tau).
\label{eq:orthonormal}
\end{equation}
 These appear not to have been used before in wave systems in our context, however they are the method of choice for obtaining decaying solutions to the Vlasov--Maxwell equation  \cite{holloway1996spectral,schumer1998vlasov} 
  in plasma physics where the setting is rather different. Nonetheless, we can draw upon the plasma physics literature and note that  
  the orthonormal Hermite functions are numerically well conditioned \cite{boyd2017dynamics} and, although in principle their efficiency is influenced by the choice of the scaling, $\tau$, and shifting $\Delta$, parameters \cite{boyd1980rate,BOYD1984382}  (see section 3.1), in practice we can simply take many modes.
  Using recurrence relations, given in Appendix A, and the orthogonality of the basis functions, Eq. \eqref{eq:KLBloch} is reformulated as:

\begin{align} \label{eq:Full}
\begin{split}
\sum\limits_{n,m}\left(\alpha\delta_{p,m-4} + \beta\delta_{p,m-2} + (\gamma-\Omega^2)\delta_{pm} + \xi\delta_{p,m+2} + \lambda\delta_{p,m+4} \right)\delta_{nn'}W_{nm} \\
=\frac{\Omega^2}{{a}}\sum\limits_{j=1}^{N_{e}}M^{(j)}\exp[{-i(G_{n'}-\kappa)x_{\mathbb{1}}^{(j)}}]\psi_{p}((y_{\mathbb{1}}^{(j)}-\Delta)/\tau)w_{\mathbb{1}}^{(j)}, 
\end{split}
\end{align}
 for integer $n^\prime$, 
where $G_{n'}$ are reciprocal lattice vectors that arise from utilising the orthogonality condition: Appendix A details the full expressions of the coefficients. 
A generalised eigenvalue problem, 
\((\matA - \Omega^2\matB)\mathbf{W} = 0\), 
 is derived by writing \eqref{eq:Full} in matrix form. In practical terms we must truncate the infinite sums and take $\mathcal{N}$ Fourier modes and $\mathcal{M}$ Hermite modes, such that $n = -\mathcal{N}, \hdots , \mathcal{N}$ and $m = 0, \hdots , \mathcal{M}$.
 The vector of unknowns, \(\mathbf{W}\), is
\begin{align}
\begin{split}
\mathbf{W} = [W_{-\mathcal{N}0},&\hdots,W_{-\mathcal{N}\mathcal{M}},W_{-\mathcal{N}+1,0},\hdots,W_{-\mathcal{N}+1,\mathcal{M}},\hdots, \\ &\hdots,W_{\mathcal{N}0},\hdots,W_{\mathcal{N}\mathcal{M}},w_\mathbb{1}^{(1)},\hdots,w_{\mathbb{1}}^{(N_{e})}]^T,
\end{split}
\end{align}
with the matrices $A$ and $B$ as
\begin{align}
\begin{split}
\matA = \begin{pmatrix}
\boxed{\mathbf{K}_{n}}_{\mathcal{M}\times \mathcal{M}} &Z_{(\mathcal{N}\times \mathcal{M})\times N_{e}} \\ 
P_{N_{e}\times(\mathcal{N}\times \mathcal{M})} & -\mathbb{1}_{N_{e}\times N_{e}}
\end{pmatrix}, \\ \\
\matB = \begin{pmatrix}
\boxed{\mathbb{1}_{n}}_{\mathcal{M}\times \mathcal{M}} & \frac{M^{(j)}}{a}P^{\dagger}_{(\mathcal{N}\times \mathcal{M})\times N_{e}}\\
Z_{N_{e}\times(\mathcal{N}\times \mathcal{M})} & Z_{N_{e}\times N_{e}}
\end{pmatrix}.
\end{split}
\label{eq:AandB}
\end{align}
The block-diagonal form of \(\mathbf{K}_n\) is helpful to note

\begin{align}
\begin{split}
\mathbf{K}_{n} = \begin{bmatrix}
\gamma\delta_{p,m}&0&\xi\delta_{p,m+2}&0& \lambda\delta_{p,m+4}&0 \\
0&\gamma\delta_{p,m}& 0	& 	\xi\delta_{p,m+2}	& 	0		& 	\ddots		    \\ 
\beta\delta_{p,m-2}& 0& \gamma\delta_{p,m}&  0	&  \xi\delta_{p,m+2}  	  	&    \ddots		  	\\
0 & \beta\delta_{p,m-2}& 0 &\gamma\delta_{p,m} &  0 	&  \ddots   		    \\
\alpha\delta_{p,m-4}&0 &\beta\delta_{p,m-2} & 0 	& 	\gamma\delta_{p,m}		& 	\ddots		 \\
0 	&\ddots& \ddots	&	\ddots& \ddots	& \ddots   
\end{bmatrix}
\end{split}
\end{align}
 as also are the explicit forms of the coefficients given in Appendix A. 
In \eqref{eq:AandB}  \(Z\) denotes a zero matrix, and \(P\) is  a matrix whose \(j^{th}\) row is 

\begin{align}
\begin{split}
P^{(j)} = \bigg[&e^{i(G_{-\mathcal{N}}-\kappa)x_{\mathbb{1}}^{(j)}}\psi_{0}((y_{\mathbb{1}}^{(j)}-\Delta)/\tau),\hdots,  e^{i(G_{-\mathcal{N}}-\kappa)x_{\mathbb{1}}^{(j)}}\psi_{\mathcal{M}}((y_{\mathbb{1}}^{(j)}-\Delta)/\tau) , \hdots,\\ &e^{i(G_{\mathcal{N}}-\kappa)x_{\mathbb{1}}^{(j)}}\psi_{0}((y_{\mathbb{1}}^{(j)}-\Delta)/\tau),\hdots ,e^{i(G_{\mathcal{N}}-\kappa)x_{\mathbb{1}}^{(j)}}\psi_{\mathcal{M}}((y_{\mathbb{1}}^{(j)}-\Delta)/\tau)\bigg],
\end{split}
\end{align}
and \(P^{\dagger}\) is the conjugate transpose of $P$.
 This eigenvalue problem is straightforward to implement numerically, we do so in Matlab, to quickly obtain accurate dispersion curves.

\section{Arrays and Interfacial Edge states} \label{RB}
We now present results for three cases of interest. We begin by considering RB waves, which are not ideally suited to the methodology and we use this example to highlight potential issues that can arise regarding the scale factor, and also generalise the high-frequency homogenisation of \cite{antonakakis12a} for doubly periodic plates to these line arrays. We then turn to graded arrays of ring-shaped arrangements of masses to illustrate how the methodology is applied to a design problem, here motivated by exploring hybridisation of RB waves along the array into modes in the bulk or trapped modes. Finally, we briefly consider a topical model using topological arangements of masses from valleytronics \cite{Pal2017,Makwana2018a}. 

\subsection{Rayleigh--Bloch modes}
\begin{figure}
    \centerline{
    \begin{tabular}{cc}
        (a) & (b) \\
      \hspace{1.25cm} \includegraphics[scale = 0.3]{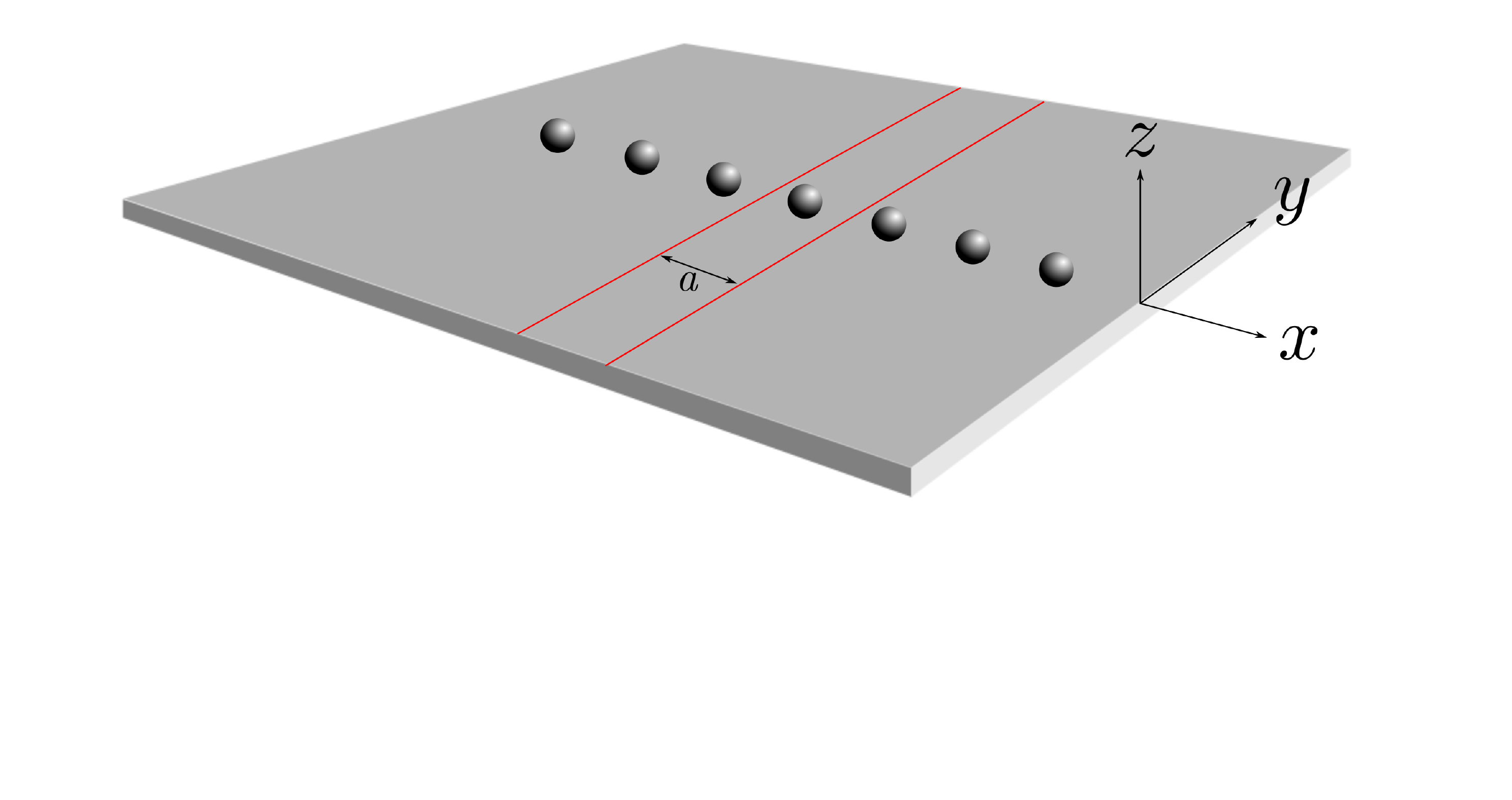} & \hspace{-1.25cm}\includegraphics[scale = 0.3]{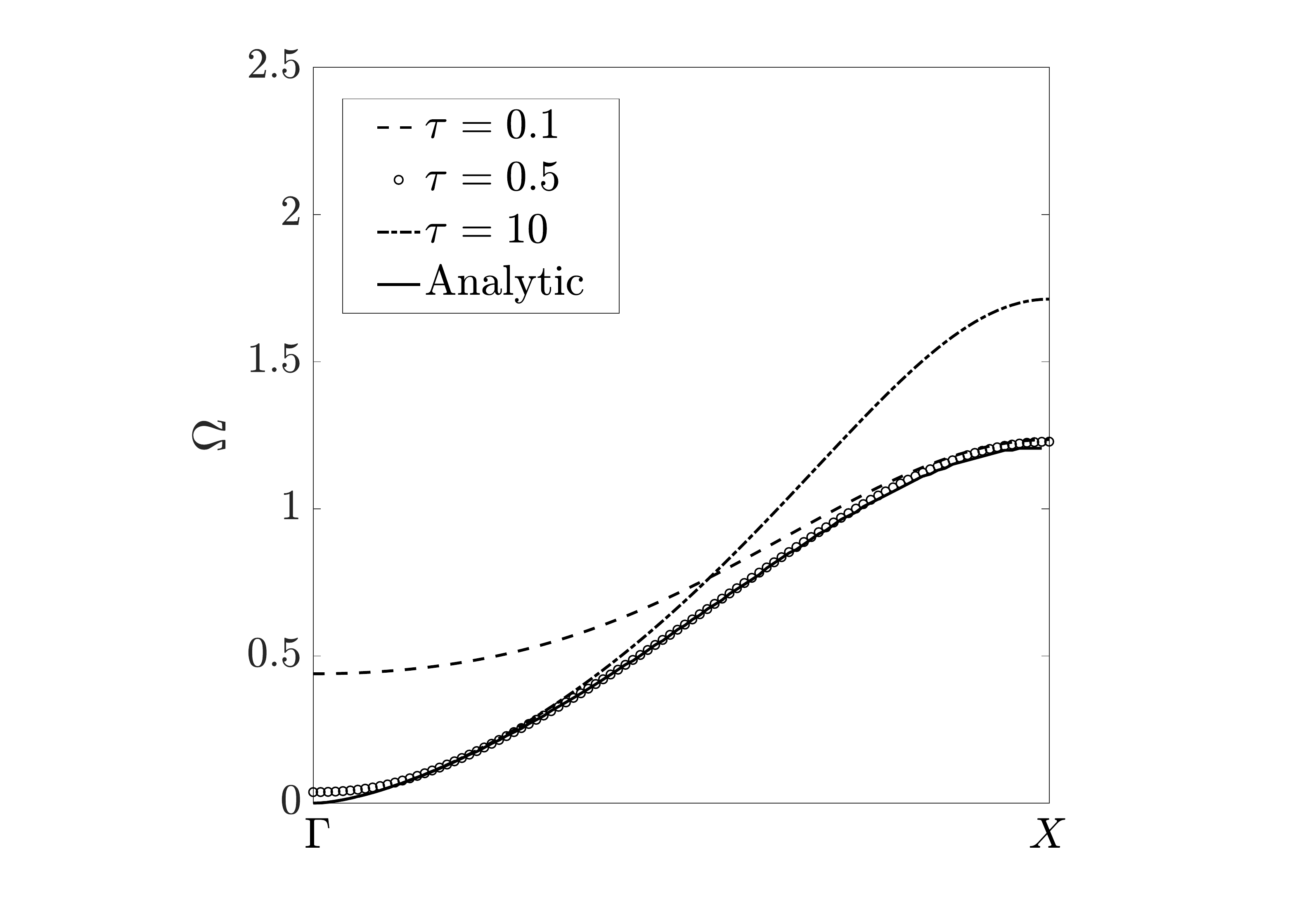} 
    \end{tabular}
    }
    \caption{(a) A schematic of an elastic plate with the fundamental strip indicated by the dotted lines and the point masses represented by spheres. (b) The effect of differing scale factors on the dispersion curves for a mass value of $M = 40$, also shown is the curve from the analytic result of \cite{evans2007penetration}. Three Fourier modes and 100 Hermite modes were used in the truncated summations, with a strip width of $a = 1$. On the horizontal scale we use the standard wavenumber notation with $\Gamma (\equiv \kappa=0)$ and $ X (\equiv \kappa=\pi/2a)$.}
    \label{fig:Fig2}
\end{figure}

\hspace*{\fill} \\
RB waves are characterised by their localisation to a structured array; they propagate along the array and decay perpendicular to it \cite{porter2005embedded} and would not exist in the absence of the array. We consider the simplest geometry for which they exist for flexural waves on an elastic plate as shown in Figure~\ref{fig:Fig2}(a), that of a one-dimensional infinite array of point masses on an infinite, thin elastic plate.

\begin{figure}[ht!]
\centering
\makebox[\textwidth][c]{\includegraphics[scale = 0.125]{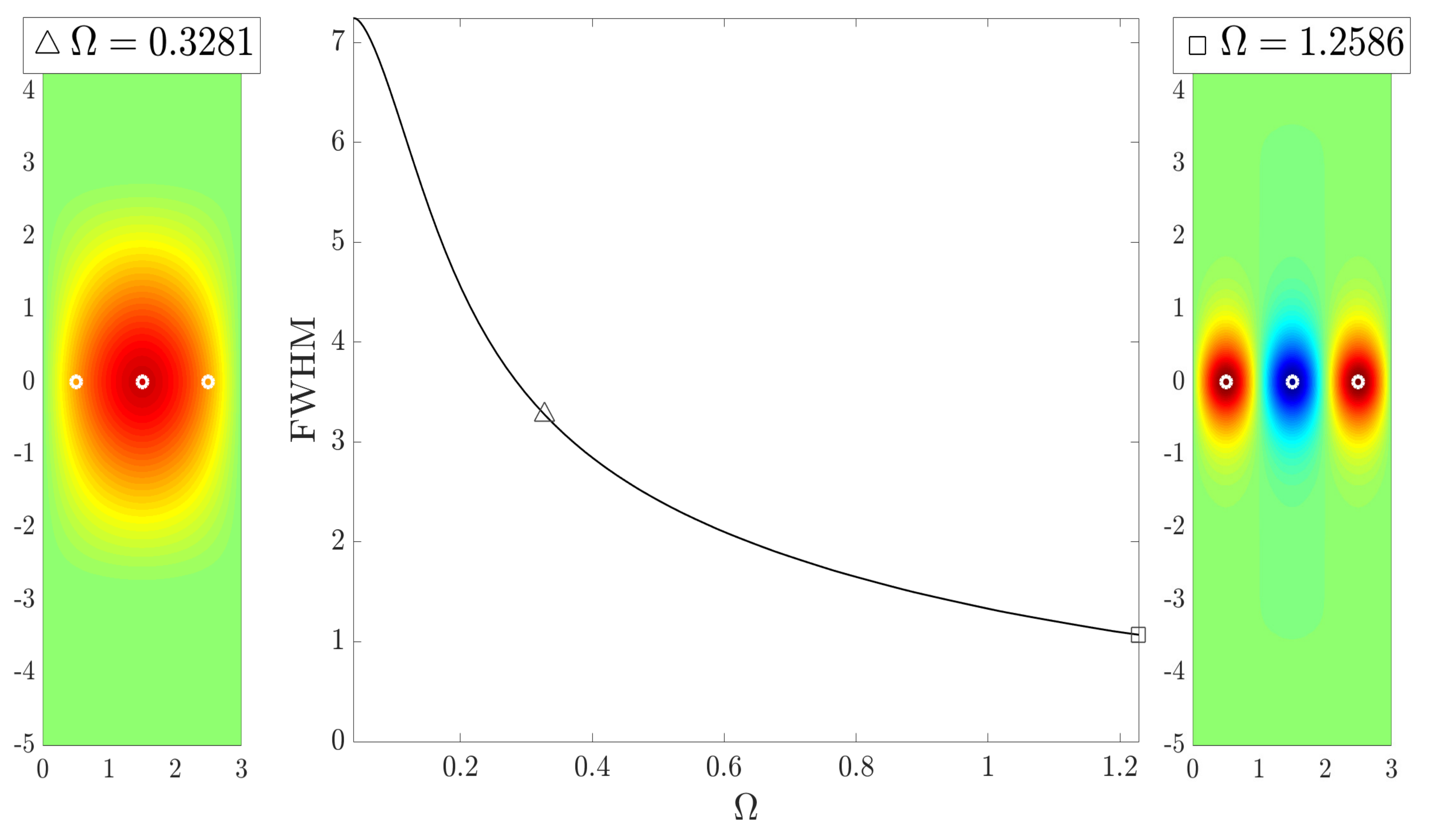}}
\caption{The variation of the full wave half-maximum (FWHM) as a function of frequency for the one-dimensional array of point masses (same parameters as in Figure~\ref{fig:Fig2}(b), $\tau=0.5$). Also shown is the wave field for three fundamental cells; the variation in decay is clearly seen.} 
\label{fig:Fig3}
\end{figure}

This system is thoroughly explored in \cite{evans2007penetration}, permits an analytical dispersion relation, and neatly provides a point of comparison for the spectral method. The numerical scheme yields dispersion curves,  shown in Figure~\ref{fig:Fig2}(b), which match the analytical curve, provided an appropriate scale factor \(\tau\) is used. Figure~\ref{fig:Fig2}(b) shows the effect of changing the scale factor for a fixed number of modes. It is possible to get a substantial mismatch between analytical and numerical results, and this arises from the differing forms of decay of the RB waves and the Hermite functions; the latter can decay more quickly than the RB mode and thus leading to inaccuracy. An optimisation problem then exists, at least for slowly decaying RB waves, between either calculating a suitable scale factor or increasing the number of modes taken in the computation. In the allied literature on Hermite expansions, \cite{tanguy1995optimum,e1996determination,den1998optimal,tanguy1999improved}, 
 there have been several attempts to optimise this process, all of which rely on knowing \textit{a priori} the form of the decay, and so in general, without an analytical solution, this is not a straightforward endeavor. However, utilising the sparsity of the matrices involved, the many zeros in the matrices can be squeezed out of the numerical scheme; this allows many modes to be used with little computational cost until convergence is observed. Nonetheless, it is worthwhile keeping the observations of Figure~\ref{fig:Fig2}(b) in mind to prevent a naive application of the method. The variation in the decay of the RB waves, for differing frequencies, is quantified by the full wave half-maximum (FWHM) of the wave-field, shown in Figure~\ref{fig:Fig3} and one can see that the decay varies strongly with frequency. In practical terms, characterising and predicting this decay has applications in assessing the coupling capabilities between line arrays and, for example, structured ring resonators \cite{mastersthesis} for acoustic ring-resonator devices and filters. 


\begin{figure}
\centering
\includegraphics[scale=0.25]{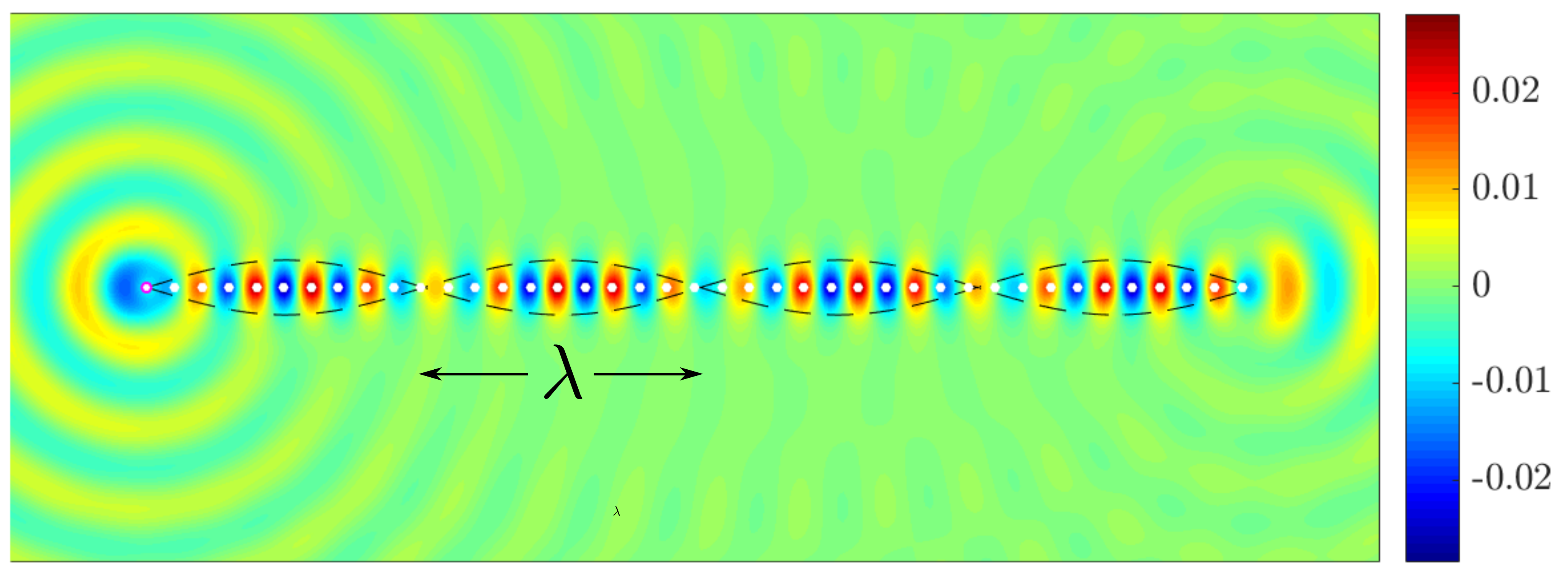}
\caption{Real part of the field shown for a scattering simulation with $M = 10$ and 41 masses forming the array, excited with a point source located on the left-most mass at frequency $\Omega = 2.15$. Also shown is the HFH wavelength from the asymptotics one obtains  $\lambda = 10.1069$ ($|T| = 2.9264$, $\Omega_{0} = 2.398$).}
\label{fig:Fig4}
\end{figure}

For line arrays, and edge modes, 
 the behaviour near the band edge, i.e. at the point $X$ in the dispersion curves, is important as simulations of a forced array show a longer wavelength envelope as well as the short-scale Bloch behaviour, see Figure \ref{fig:Fig4}, which reflects the inherent two-scales of wave propagation in these systems. This wavelength, together with an effective equation for the long-scale behaviour, follows from two-scale asymptotic methods \cite{bensoussan78a}, in particular using high frequency homogenisation (HFH)  \cite{antonakakis12a,Craster2010}. Here we adapt the methodology to
  treat the line array and develop the procedure directly in reciprocal space, the details are given in \ref{sec:HFHAppend}, and this is particularly useful given the solution method we have developed. The basic idea is to split the spatial coordinate into two scales to be treated independently: a short scale defined by the coordinates $\boldsymbol{\xi} = (x/l,y/l)$, and a long scale $X = x/L$ such that $\epsilon \equiv l/L \ll 1$ where $l$ and $L$ are typical length-scales in the long- and short-space. By asymptotically expanding the wave field and frequency obtained through the spectral method, an effective medium equation is derived in the form of an ordinary differential equation (ODE) for the long scale modulated by oscillatory behaviour on the short scale. The upshot is that the field $w({\bf x})=f(X) W_0 (\boldsymbol{\xi})$ 
where $W_0$ is the Bloch eigensolution at the band-edge and $f(X)$ is a long-scale envelope function which is found from an effective ``string" equation
\begin{align}
f_{XX} - \left(\frac{\Omega^2-\Omega_{0}^2}{T}\right)f = 0.
\label{eq:fequation}
\end{align}
Here $\Omega_{0}$ is the frequency obtained at the band edge and $T$ is a geometry dependent parameter that is calculated from the spectral method or directly from the eigenvalue problem, and related analysis, as in \ref{sec:HFHAppend}. 
The long-scale envelope is observed for excitation frequencies close to the band edge with a characteristic wavelength defined by
\begin{align}
\lambda = 2\pi\sqrt{\frac{T}{\left(\Omega^2-\Omega_{0}^2\right)}},
\end{align}
and this is seen in scattering simulations, see Figure~\ref{fig:Fig4}. 
This direct scattering simulation, uses the well-known Green's function for the KL equation, 
 to construct a linear system, as in \cite{Torrent2013,Makwana2018a,evans2007penetration}, that is solved using standard methods. 

The Fourier-Hermite method clearly captures the analytical results for the one-dimensional infinite array of point masses, and dovetails nicely with homogenisation to produce an effective equation valid for frequencies close to the band-edge. This RB example is, 
due to potentially large differences in the decay rate of the RB waves and the Hermite functions, one of the least optimal geometries for the implementation of this Fourier-Hermite method, and so it is reassuring to see its accuracy; this decay rate discrepancy occurs as  the RB wave passes through the sound-line at the origin. A more typical scenario, and where the real power of this method lies,  is for finding dispersion diagrams for more strongly localised modes, such as those confined within more exotic geometries of line arrays, such as the array of rings in the following subsection, and those which are created by topology. For those, the Hermite functions are known to oscillate between the turning points, $\psi_t$, where $|\psi_t| \approx \tau\sqrt{2m+1} + \Delta$, and after which they monotonically exponentially decay, estimating where this decay is expected gives scaling and shifting parameters. Coupling this with the ability to utilize the sparsity structure  of the matrices involved, results in a fast, accurate method for finding the dispersion curves of decaying waves. To demonstrate the robustness of the Fourier-Hermite method we now implement it for more complicated geometries of line arrays. 

\subsection{Graded Ring Arrays}
\begin{figure}[!ht]
\centering
\includegraphics[scale = 0.3]{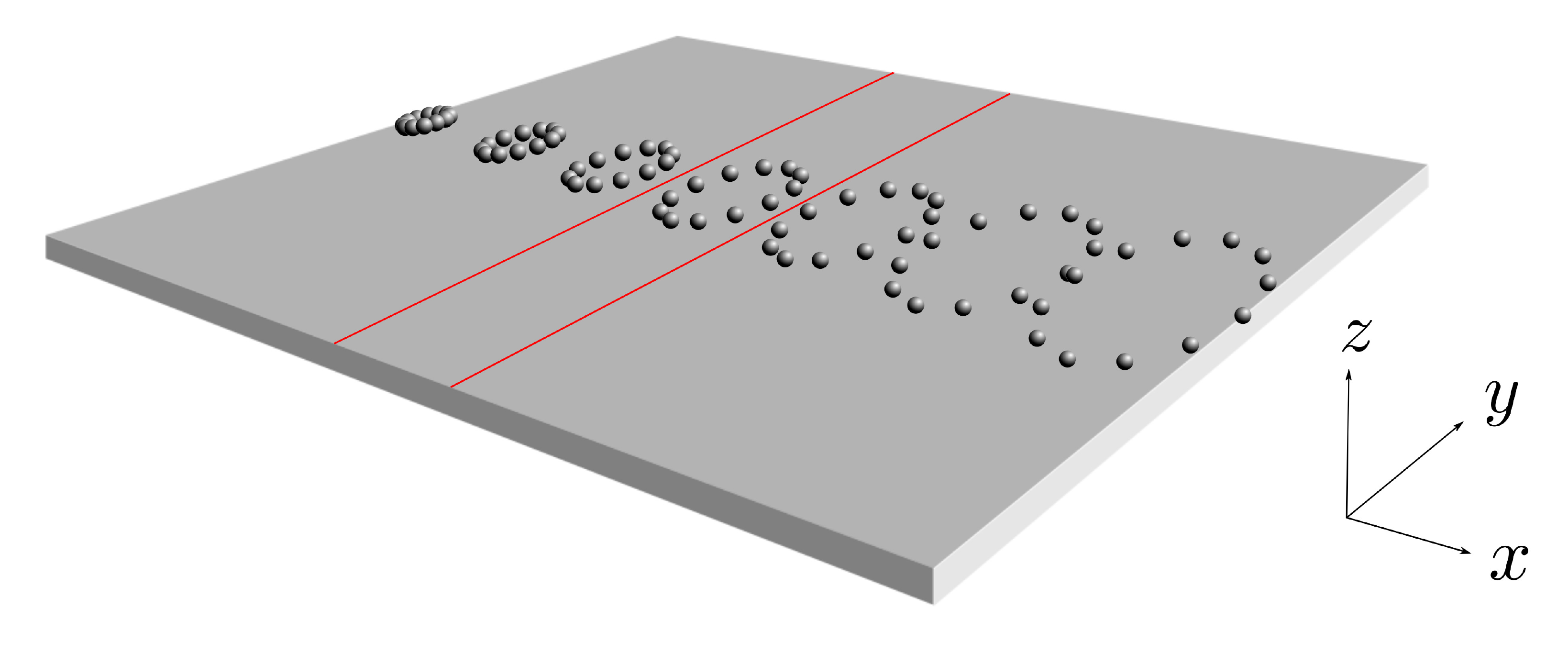}
\caption{Schematic of ring resonators forming graded line array, with unit strip of infinite medium shown in red.}
\label{fig:Fig5}
\end{figure}

The efficiency of the Fourier-Hermite method allows for parametric studies involving complex geometries of scatterers to be quickly investigated. We consider graded line arrays of point masses arranged around in a ring, see Figure~\ref{fig:Fig5} as this has interesting properties due to the symmetries of the eigenmodes. Grading the array leads to rainbow trapping which is the spatial separation of constituent frequencies of a source, so that the band-gap cut-off disallowing wave propagation is experienced at different positions along the array. This effect has been realised in optics for some time \cite{Tsakmakidis2007,Gan} and subsequently in acoustics \cite{Romero-Garcia2013,Zhu2013}, seismology \cite{Colombi2016} and water waves \cite{bennetts18a} using single scatterers in each fundamental cell. For the ring of point masses chosen here (we use 16 masses in the ring) the Fourier-Hermite scheme rapidly gives the dispersion curves for different ring radii, and we identify these as candidates to perform the rainbow trapping effect for RB flexural waves, without exploiting resonance effects; typically resonances are used, i.e. \cite{Colombi2016}, to push the effect down in frequency so that  sub-wavelength devices are designed, but, as here, resonance is not essential for the effect. We achieve the trapping and hybridisation effects by employing the coupling properties, or lack thereof, between antisymmetric (odd) and symmetric (even) modes supported by the array.

\begin{figure}[t]
\centering
\includegraphics[scale = 0.06]{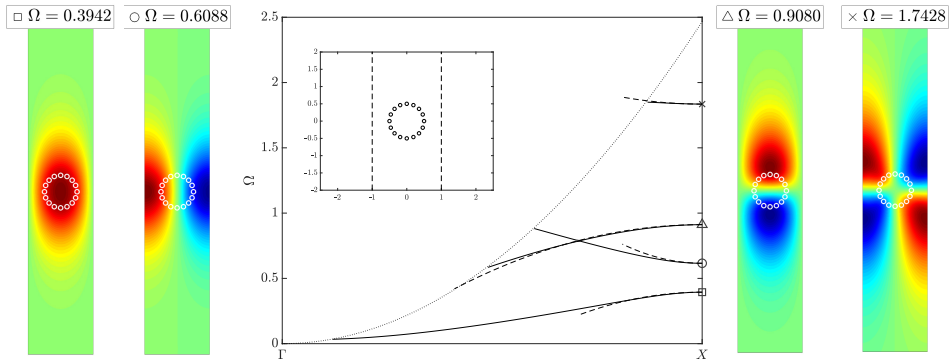}
\caption{Dispersion curves for arrangements of masses placed around a ring of radius $R = 0.5$ with 16 masses of mass $M = 10$. Also shown is the wave field within a single strip at the indicated frequencies; note that the second and third modes have opposite reflectional symmetries in $x$ and $y$. The dashed lines show the asymptotics obtained from the HFH method.}
\label{fig:Fig6}
\end{figure}

Figure~\ref{fig:Fig6} shows the dispersion curves for a typical radius value, $r=0.5$, along with the field plots at the first four frequencies along the wavenumber associated with $X$, the modes anti-periodic across a cell. Also shown in Figure 6 is the sound-curve corresponding to the dispersion relation in the absence of any scatterers. There are both odd and even solutions with respect to reflection in $x$ (or $y$)  about the centre of the line array. Looking at the second and third modes in particular we note that they differ in reflectional symmetry and therefore cannot couple, it is therefore possible for their dispersion curves to cross and for an accidental degeneracy (a crossing) to occur; indeed this is the band-crossing seen in Figure~\ref{fig:Fig6}. What we observe is that by decreasing the ring radius we can push this crossing towards, and ultimately beyond, the sound-line to create a band-gap and excluded frequencies from propagation; RB modes do not occur above the sound-curve unless they are embedded \cite{porter2005embedded} and we do not consider such cases here. Moreover, the bandgap  creates a clear separation in the propagating regions for the odd and even modes, and this is now used to produced the trapping and hybridising of array modes.

\begin{figure}
    \centering
    \includegraphics[scale = 0.35]{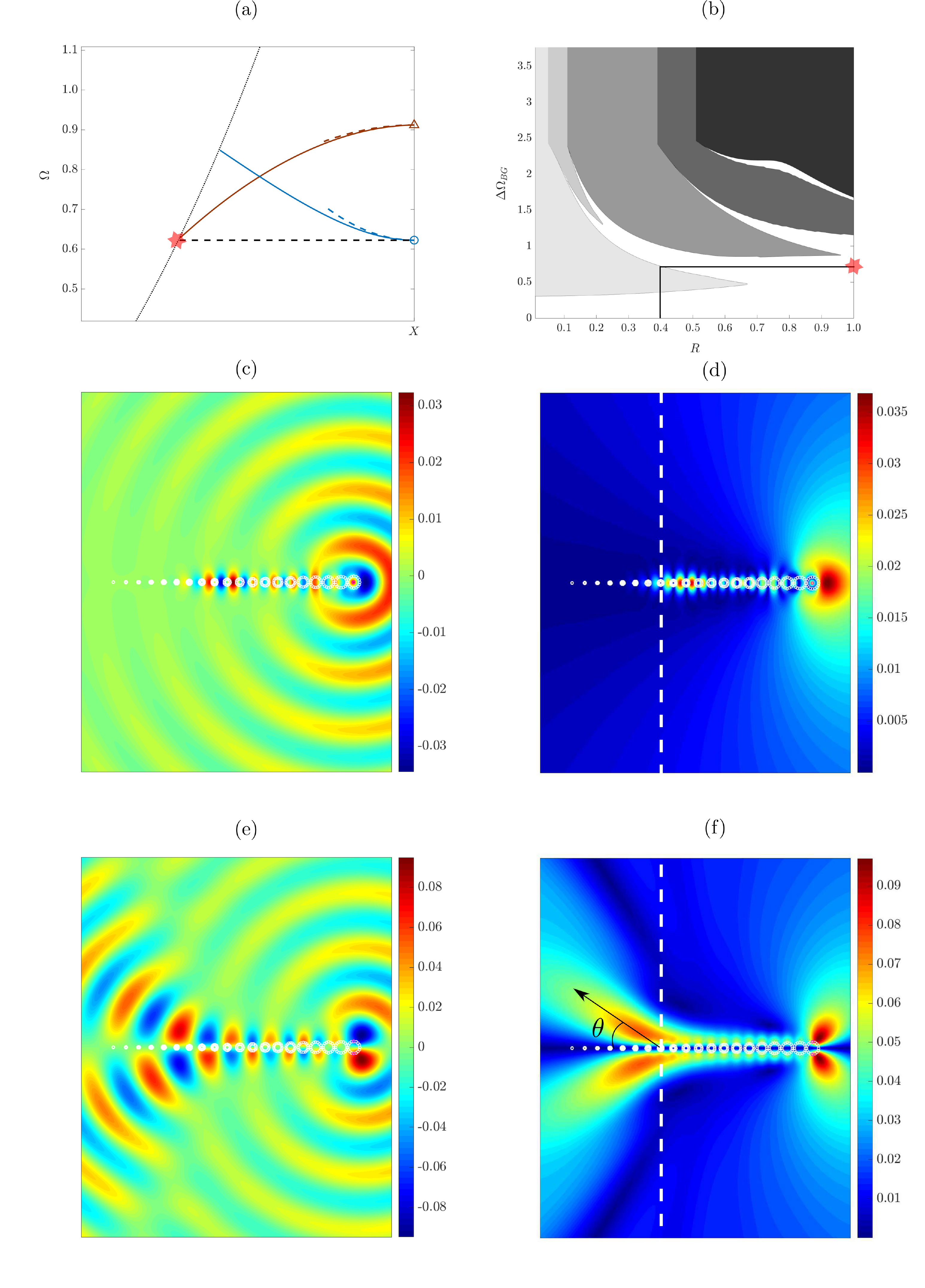}
    \caption{(a) Expanded view of dispersion curves for ring of $R = 0.4$. The excitation frequency used in (c-f) is shown as the dashed line.  (b) Propagation plot for varying ring radius, with $M = 10, ~a = 2$. Shown in grey are the band-gaps, whose width is a function of ring radius. The red star marks the excitation frequency, $\Omega = 0.7226$,  with the solid black line showing the radius for which propagation occurs. (c-d) Rainbow trapping, real and absolute value of the even mode excited by a monopole at the center of rightmost ring. (e-f) Hybridisation, real and absolute value for the odd mode excited by a dipole source. The vertical white dashed line in (d,f) indicates the spatial position where $R=0.4$.}
    \label{fig:Fig7}
\end{figure}

The Fourier-Hermite method allows us to quickly obtain the dispersion curves for many different ring radii and create a band gap diagram at each ring radius, see  Figure~\ref{fig:Fig7}(b). This allows for graded structures, as shown in the schematic in  Figure~\ref{fig:Fig5}, to be designed and their properties assessed easily.

The band structure for a linearly graded ring radius is shown in Figure~\ref{fig:Fig7}(b), where the shaded regions show the band gaps. The knowledge of whether each of the eigensolutions is odd or even determines, for a particular frequency, where and whether rainbow trapping or hybridisation with a propagating plate mode occurs. In this `propagation plot', the solid black line shows the frequency of excitation of monopole (dipole) sources used to generate even (odd) RB waves as shown in Figures~\ref{fig:Fig7}(c-f). For larger radii there is no band gap at this frequency and both modes are supported. At a radius of $0.4$ a (purely geometrical) local band gap exists at this frequency. An expanded view of this section of the dispersion curve is shown in Figure~\ref{fig:Fig7}(a), where the curves have been coloured blue (even) and red (odd) and the HFH asymptotics are also shown and the excitation frequency is shown with a star and dashed line. For subsequently smaller radii, these curves are pushed to higher frequencies and so this excitation frequency lies in the band-gap for both modes. At the spatial position corresponding to the cut-off radius ($R=0.4$), for the odd-mode the dispersion curve meets the light line whilst for the even mode its dispersion curve is at the band-edge (X); these modes cannot couple into each other due to their opposing symmetry. The odd and even modes therefore behave differently, the even mode is trapped and reflected back along the array, since it cannot couple into the odd mode, and its at a band-edge and its energy has nothing to couple into; this is  the rainbow trapping effect, albeit not due to resonance effects, and is shown in Figures~\ref{fig:Fig7}(c-d). At the same position, the odd mode  encounters  the light line and its energy can couple into the bulk. Due to the grading of the array, this mode cannot continue to propagate along the array and instead hybridises with a plate wave, as shown  in Figures~\ref{fig:Fig7}(e-f). The angle at which this occurs matches the Snell's law prediction of $\theta = \sin^{-1}\left(c_{RB}/c_p(\omega)\right) \simeq 34^{\circ}$, where $c_{RB}, c_p$ are the RB and plate wave speeds respectively.
The radii at which these effects occur matches well the predictions from the dispersion relations, as shown by the vertical white dashed line in Figures~\ref{fig:Fig7}(d,f).  Therefore one can obtain very different phenomena just by simply switching the source from a monopole to a dipole to induce the different symmetries.

\begin{figure}[ht]
\centering
\includegraphics[scale = 0.375]{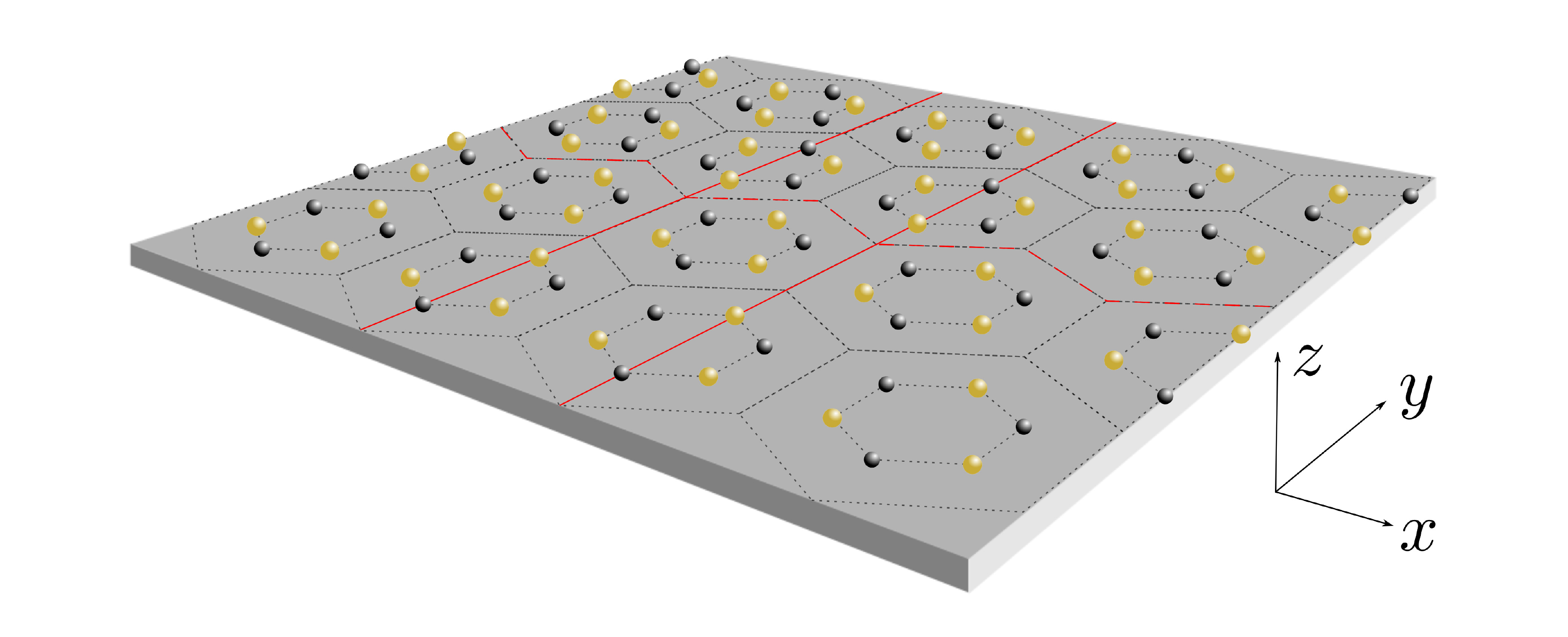}
\caption{Schematic of the zig-zag interface (dashed red) between two media created by alternating masses in a hexagonal lattice geometry with the unit strip shown by thick red lines.}
\label{fig:Fig8} 
\end{figure}

 \begin{figure}
     \centering
\centerline{\begin{tabular}{cc}
(a) & (b) \\
     \includegraphics[scale = 0.2]{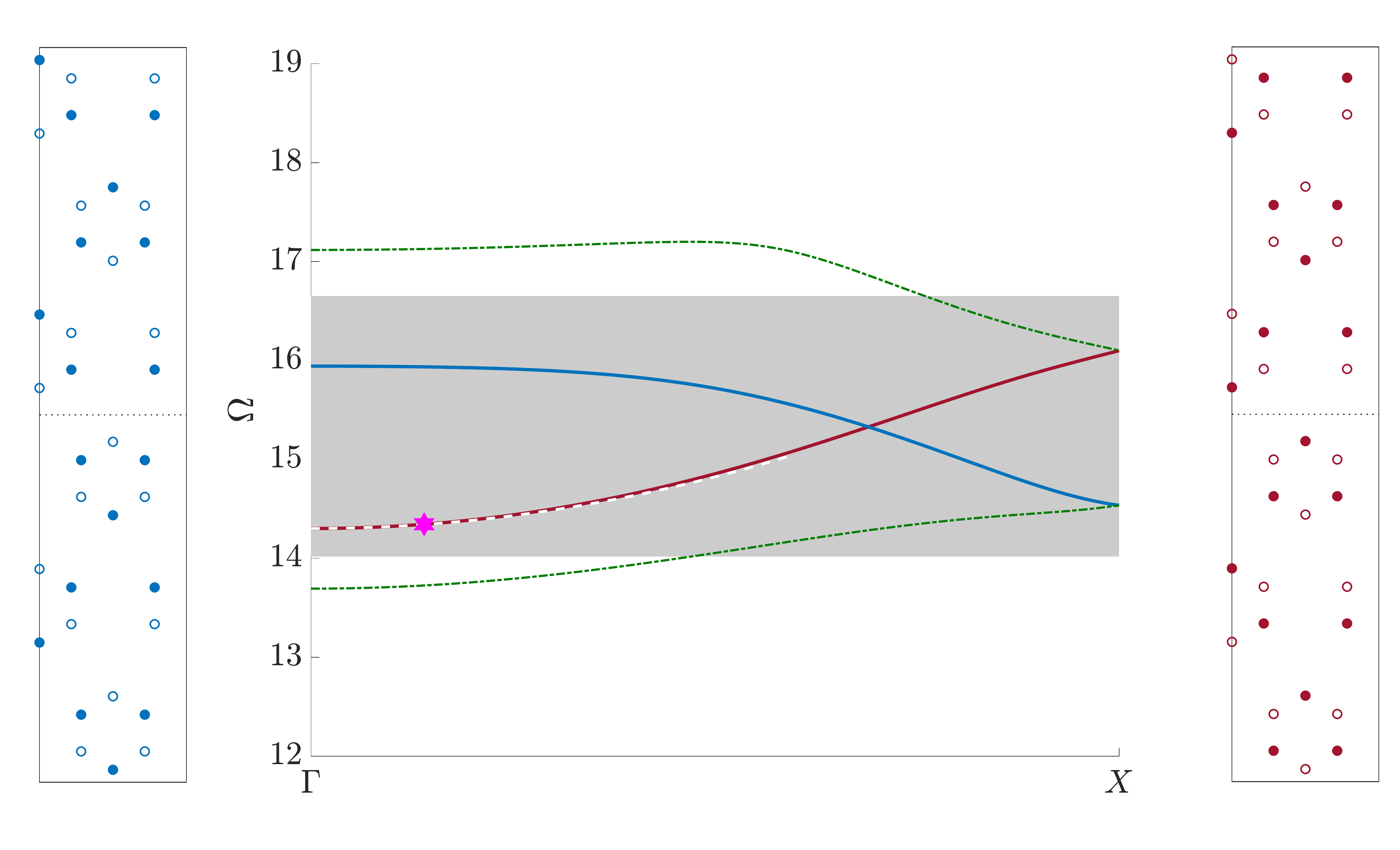}&\includegraphics[scale= 0.25]{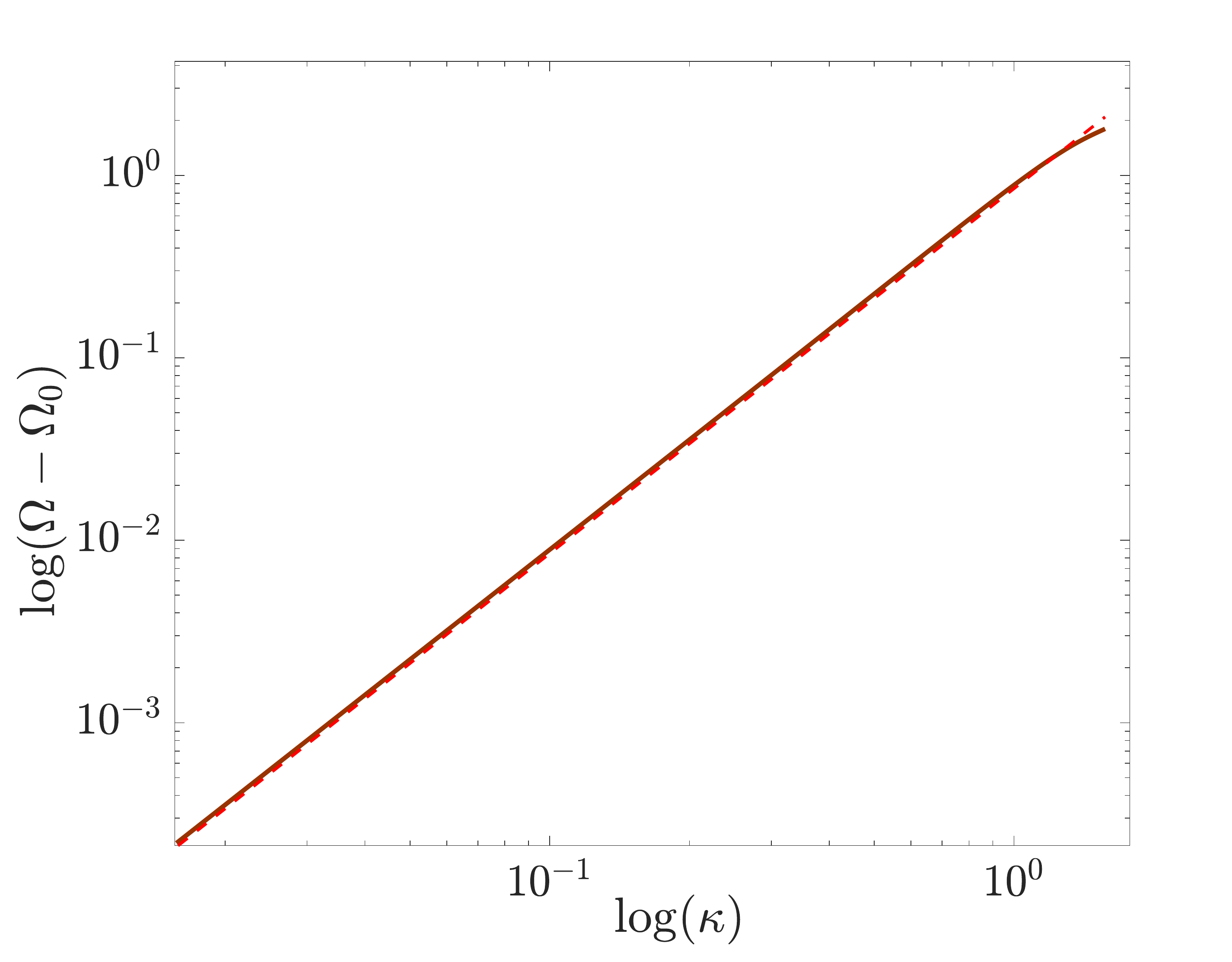}  \\ (c) & (d) \\
     \includegraphics[scale = 0.25]{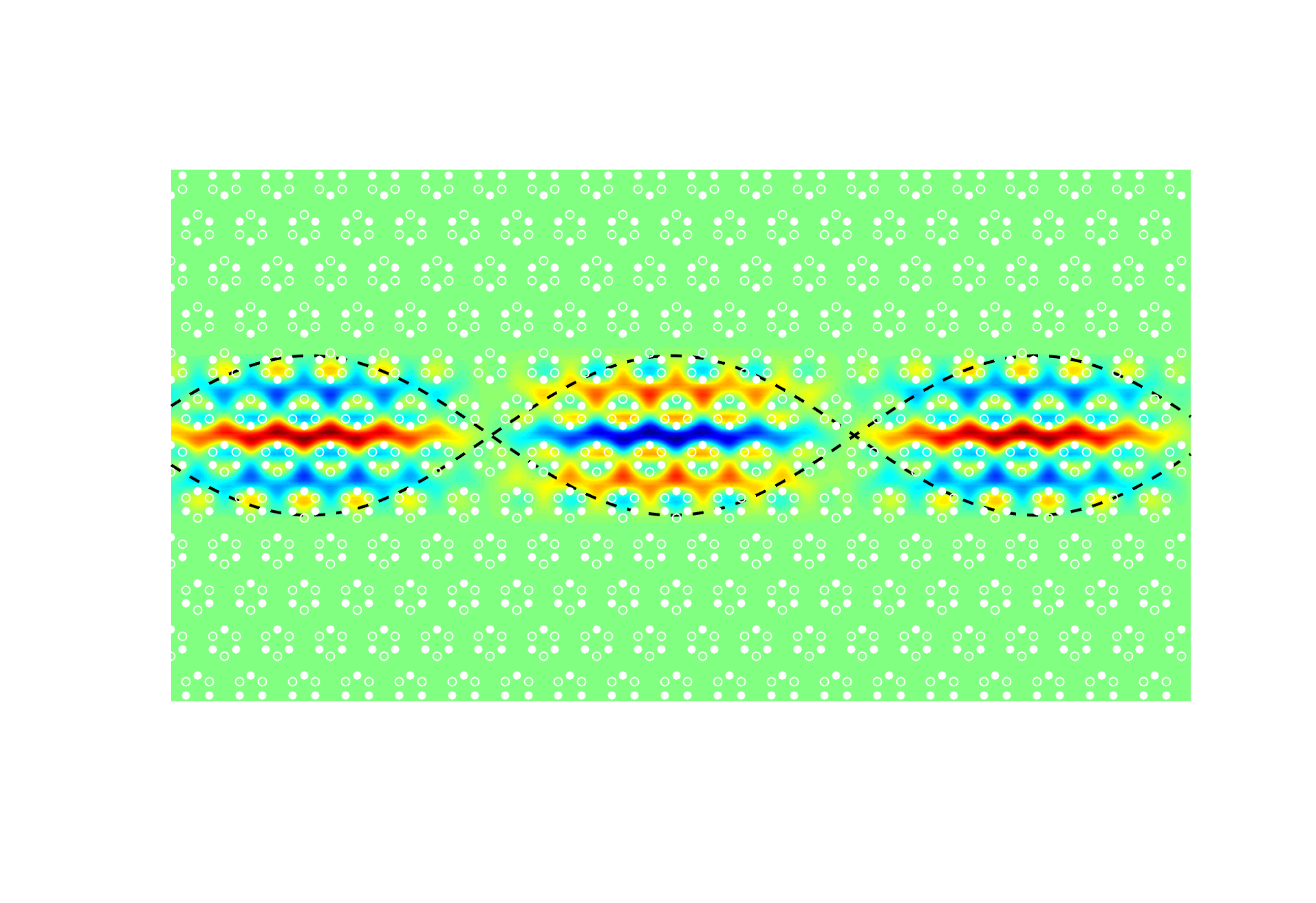}&\includegraphics[scale = 0.1]{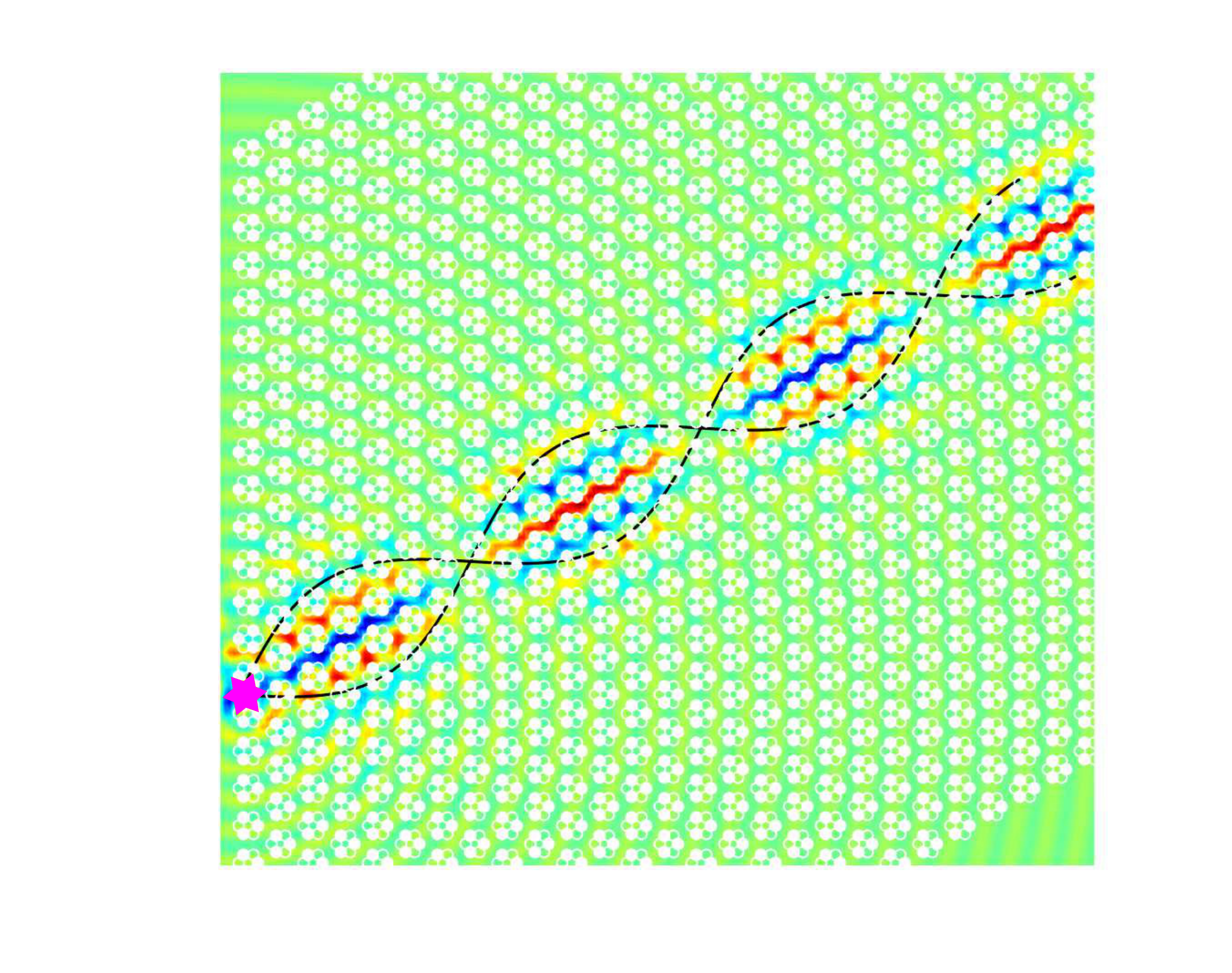} 
\end{tabular}}
    \caption{(a) Dispersion diagram for edge states, located within the band gap of the infinite medium. The mass distribution is such that the hollow circles represent a mass of $M_{1} = 1$ and the filled circles $M_{2} = 4$. The strips show the crystal arrangements that give the dispersion curves of the corresponding colour. The dashed green curves show other leaky decaying solutions which can couple to propagating modes within the bulk (b) Convergence of asymptotics for the red curve and the HFH curve (dashed red line) obtained from the matrix HFH. (c) Predicted eigensolution from method at frequency $\Omega = 14.34$ (marked by magenta star in (a)). (d) Scattering solution excited at same frequency by monopolar source at position of magenta star.}
\label{fig:Fig9}
 \end{figure}
 
The method is also suited to searching for more rapidly decaying solutions than the array modes described above. Attention is now turned to the very recent topological edge states and further emphasises the utility of the scheme.

\subsection{Topological Edge States}
It has recently been shown that time-reversal symmetric (TRS) valley-Hall insulators \cite{Xiao2007}, that are ultimately topologically trivial, can be constructed in elastic wave systems \cite{Pal2017,Makwana2018a,Makwana2018b}. By attaching two geometrically distinct media, that have opposite valley Chern numbers, robust edge states arise; these states are more commonly known as zero-line modes (ZLMs).  Moreover these robust modes can propagate around corners, with limited backscattering, and be used to create wave-energy splitters \cite{Makwana2018b}. We adapt an example from \cite{Makwana2018a} and construct a hexagonal structure with six masses within each hexagonal cell; masses, that alternate between $M_1$ and $M_2$, are placed at the vertices of the inner hexagon, see Figure~\ref{fig:Fig8} and \ref{fig:Fig9}(a). We then place this medium above its $\pi/3$ rotated twin (see Fig. \ref{fig:Fig9}(a)) to yield two broadband edge states that reside within the band-gap. Alternating masses has quite profound consequences, as shown in \cite{Makwana2018a}, taking all masses identical gives a symmetry induced Dirac point in the bulk dispersion curves, and alternating masses gaps this Dirac point yielding a topologically nontrivial band-gap, shown in Figure \ref{fig:Fig9}(a). Placing one medium next to its rotated twin leads to edge states within the band-gap; that take different form depending upon whether one medium is above/below the other. These geometrically induced edge states have interesting properties; there are localized zero-displacement regions around the interface, which in turn create a self-pinned linear array, along which RB modes propagate \cite{Makwana2018b}. This particular subfield of topological insulators, that utilises the valley-Hall effect, is often referred to as valleytronics and it is a vibrant area where ideas from condensed matter physics are being applied to classical wave systems \cite{khanikaev_two-dimensional_2017}. An application of these TRS topological insulators, is to leverage their robustness to direct waves around sharp bends or demonstrate their backscattering immunity against defects; there are a myriad of applications, for example, in telecommunications \cite{shalaev18a} for the analogous electromagnetic case. Similar ambitions are evident in elastic waves with the direction, and re-direction, of mechanical vibration \cite{Pal2017,Makwana2018b}.   

The accurate identification of edge states is a vital component of these studies, and one usually takes a long finite ribbon, with the subsequent eigenstates of interest emerging within the band, but due to the truncated system there are a copious number of spurious modes. 
As shown in Figure \ref{fig:Fig9} the Fourier-Hermite methodology (and the HFH shown as the dashed curve), can be successfully applied here with only the decaying states detected; this makes this approach numerically advantageous. The importance of the periodicity of the energy-carrying envelope for propagation around bends was emphasised in \cite{Makwana2018b}; hence the ability to mathematically characterise this envelope, via HFH, is incredibly useful. A pictorial demonstration of this is shown in Figure \ref{fig:Fig9}(c,d); where the envelope was obtained by applying the HFH methodology in the vicinity of $\Gamma$. 
The displacements, and envelopes, shown in Figure \ref{fig:Fig9}(c, d) are derived from the eigenstates and scattering simulation, respectively. 

\section{Discussion and Concluding Remarks}
\label{Conc}

It is worthwhile discussing the tactical choice of expansion functions taken here in the context of an alternative. Here we chose to exploit a Fourier-Hermite basis and other expansions are possible, the Fourier expansion is, of course, natural for the quasi-periodic Bloch boundary conditions, but capturing the exponential decay can be done spectrally in, at least, two different ways. The main alternative to using the Hermite expansion employed herein, as the decaying basis, is to use a Laguerre basis instead. Mathematically, there are many similarities between the current problem and that of extracting trapped and localised wave solutions in deformed waveguiding systems \cite{postnova2008trapped}; there spectral methods based upon Laguerre basis expansions automatically build in the exponential decay and Chebyshev expansions were adopted across the waveguide. Those approaches were attractive as they were based upon spectral collocation \cite{trefethen00a} in the physical domain and thus gave spectrally accurate solutions; furthermore  since relatively few points were required they also led to the rapid identification of solutions. The adaption of this to the infinite ribbon eigensolution with Floquet-Bloch side conditions is not however straightforward, particularly as we aim to develop a spectral Galerkin method \cite{boyd2001chebyshev} (and not collocation) to take advantage of reciprocal space to obtain the dispersion relations that extract only the decaying modes. The Laguerre expansions are useful for solutions either even or odd around a symmetry axis and the natural domain of orthogonality for Laguerre is over $[0,\infty)$. The pure exponential decay of the Laguerre expansion functions would appear to be an advantage, whereas the Hermite have faster, Gaussian, decay, however both ultimately require scale factors. The Laguerre expansions are inconvenient when  wavefields that could be either  even or odd arise, as is generally the case, and we wish to retain that generality. In this light, we continue with the Hermite functions, and through the scaling factors we can compensate for their more rapid decay. It is none the less worth pointing out that other spectrally accurate schemes incorporating decay, and/or an infinite domain, are possible and indeed worthwhile to consider. 

Having the Fourier-Hermite methodology at our disposal then leads to its use as a design tool. The graded array and topological examples have been chosen as they are, in their own right, of interest in topical applications - energy harvesting and energy transport around sharp defects respectively. In both cases the details of the scatterers, their geometric arrangement or orientations are key to the physical effects desired. For instance for the graded array the effect we are interested in occurs over a relatively narrow frequency range, but this can be tuned by, for instance altering the pitch of the array, that is the 
 width of strip, with narrower strips shifting band gaps to higher frequencies (not shown herein) and widening them. We have concentrated upon mass-loading for simplicity, but by extending the method to include resonators, the 
results move to being subwavelength and occur over broader frequency ranges. In the case chosen, the ring which is a geometrically symmetric structure, one can take advantage of a further detail which is that the Hermite functions contain alternating even/odd functions and one can search for solutions of one symmetry or the other. Thus we anticipate the Fourier-Hermite methodology, will be of value in the design and optimisation of geometries to transport and control wave energy. 

\section*{Acknowledgements}
The authors thank the UK EPSRC for their support through Programme Grant EP/L024926/1 and a Research Studentship. RVC acknowledges the support of the Leverhulme Trust. 

\appendix

\section{Hermite Functions}
\label{sec:HermiteAppend}
The orthogonality relation of the Hermite functions in equation \eqref{eq:orthonormal} is defined through the inner product 
\begin{equation}
\langle\psi_{p}(\zeta),\psi_{m}(\zeta)\rangle \equiv \int\limits_{-\infty}^{\infty}\psi_{p}(\zeta)\psi_{m}(\zeta)d\zeta = \delta_{pm},
\end{equation}
and these functions satisfy the useful recurrence relations \cite{boyd2017dynamics}, which we use freely in the text
\begin{align}
\begin{split}
\psi_{m+1}(\zeta) = \sqrt{\frac{2}{m+1}}\zeta\psi_{m}(\zeta) - \sqrt{\frac{m}{m+1}}\psi_{m-1}(\zeta) \\
\psi_m'(\zeta) =  \sqrt{\frac{m}{2}}\psi_{m-1}(\zeta) - \sqrt{\frac{m+1}{2}}\psi_{m+1}(\zeta),
\end{split}
\end{align}
 where the prime denotes the derivative. 
These lead to the coefficients in \eqref{eq:Full} that have explicit form:
\begin{align}
\begin{split}
\alpha &\equiv \frac{1}{4\tau^4}\sqrt{m(m-1)(m-2)(m-3)},  \\ \\
\beta &\equiv-\frac{1}{\tau^2}\sqrt{m(m-1)}\left(|G-\kappa|^2 + \frac{1}{\tau^2}\left(m - \frac{1}{2} \right) \right), \\ \\
\gamma &\equiv \left(|G-\kappa|^4 + \frac{2}{\tau^2}|G-\kappa|^2\left(m + \frac{1}{2} \right) +\frac{3}{4\tau^4}\left(2m^2 +2m+1 \right) \right),  \\ \\
\xi &\equiv -\frac{1}{\tau^2}\sqrt{(m+1)(m+2)}\left(|G-\kappa|^2 + \frac{1}{\tau^2}\left(m + \frac{3}{2} \right) \right),  \\ \\
\lambda &\equiv \frac{1}{4\tau^4}\sqrt{(m+1)(m+2)(m+3)(m+4)}. 
\end{split}
\end{align}

\section{HFH in reciprocal space}
\label{sec:HFHAppend}
High Frequency Homogenisation \cite{Craster2010} allows the development of a long-scale homogenised ODE which incorporates microstructuring on a length scale comparable to that of the unit cell in periodic structures. This resulting equation can be used to predict oscillatory and decaying behaviours along the array. The usual asymptotic method involves expanding in physical space, by treating the  governing equation and eigensolution in a multi-scale manner \cite{antonakakis12a}. Instead, one can efficiently calculate the critical parameter, $T$, by expanding in Fourier space from the outset using results obtained using the spectral method. This has the advantage of dovetailing with the Fourier-Hermite method to calculate the asymptotics of the dispersion curves at any high symmetry point of the irreducible Brillouin zone and giving the quantity, $T$, required for the effective ODE in \eqref{eq:fequation}. 

We present the method for expansion about the $\Gamma$ point and take $\kappa \rightarrow \epsilon\kappa$ such that $\epsilon \ll 1$. From the Fourier-Hermite method, the two matrices $\matA(\kappa), \matB(\kappa)$, are known and the following expansions are used about $\Gamma$:

\begin{align}
	\begin{split}
	\begin{tabular}{lcc}
	$\matA = \matA_{0} + \epsilon\kappa\matA_{1} + \frac{\epsilon^2\kappa^2}{2}\matA_{2} + \hdots$ , & $\matB = \matB_{0} + \epsilon\kappa\matB_{1} + \frac{\epsilon^2\kappa^2}{2}\matB_{2} + \hdots$ , \\ \\
	$\Omega^2 = \Omega_{0}^2 + \epsilon\Omega_{1}^2 + \epsilon\Omega_{2}^2 + \hdots $ , & $\mathbf{W} = \mathbf{W}_{0} + \epsilon\mathbf{W}_{1} + \epsilon^2\mathbf{W}_{2} + \hdots$ ,
	\end{tabular}
	\end{split}
\end{align}
where $\matA_{i} = \frac{\partial^i\matA}{\partial\kappa^i}\big\rvert_{\kappa = 0}$ and similarly for $\matB_{i}$. A hierarchy of equations in orders of $\epsilon$ is then deduced;

\begin{subequations}
	\begin{align}
	\begin{split}
	\mathcal{O}(\epsilon^0): \quad \left[\matA_{0}-\Omega_{0}^2\matB_{0}\right]\mathbf{W}_{0} = 0
	\end{split} \label{eq:ep0_FS}\\
	\begin{split}
	\mathcal{O}(\epsilon^1): \quad \left[\matA_{0}-\Omega_{0}^2\matB_{0}\right]\mathbf{W}_{1} + \left[\kappa\left(\matA_{1} -\Omega_{0}^2\matB_{1}\right) -\Omega_{1}^2\matB_{0}\right]\mathbf{W}_{0} = 0
	\end{split} \label{eq:ep1_FS}\\
	\begin{split}
	&\mathcal{O}(\epsilon^2): \quad \left[\matA_{0}-\Omega_{0}^2\matB_{0}\right]\mathbf{W}_{2} + \left[\kappa\left(\matA_{1} -\Omega_{0}^2\matB_{1}\right) -\Omega_{1}^2\matB_{0}\right]\mathbf{W}_{1} \\ &\qquad\qquad+ \left[\frac{\kappa^2}{2}\left(\matA_{2}-\Omega_{0}^2\matB_{2}\right)-\kappa\Omega_{1}^2\matB_{1}-\Omega_{2}^2\matB_{0}\right]\mathbf{W}_{0} = 0.
	\end{split} \label{eq:ep2_FS}
	\end{align}
\end{subequations}

A solvability condition is enforced by multiplying on the left by $\mathbf{W}_{0}^{\dagger}$ where the $\dagger$ denotes the Hermitian transpose. Applying this to the $\mathcal{O}(\epsilon^1)$ equation gives

\begin{align}
	\begin{split}
	\mathbf{W}_{0}^{\dagger}\left[\matA_{0}-\Omega_{0}^2\matB_{0}\right]\mathbf{W}_{1} = -\mathbf{W}_{0}^{\dagger}\left[\kappa\left(\matA_{1}-\Omega_{0}^2\matB_{1}\right)-\Omega_{1}^2\matB_{0}\right]\mathbf{W}_{0}.
	\end{split}
\end{align}
Providing 
\begin{equation}
\mathbf{W}_{0}^{\dagger}\left[\kappa\left(\matA_{1}-\Omega_{0}^2\matB_{1}\right)\right]\mathbf{W}_{0} = 0,
\label{eq:provided}
\end{equation}
then $\Omega_{1}^2$ is forced to vanish; this is for non-degenerate eigenvalues as we have in this article.  This allows calculation of $\mathbf{W}_{1}$ as 
\begin{equation}
\mathbf{W}_{1} = -\kappa\left[\matA_{0}-\Omega_{0}^2\matB_{0}\right]^{+}\left(\matA_{1}-\Omega_{0}^2\matB_{1}\right)\mathbf{W}_{0}.
\end{equation}
 where $+$ denotes the pseudo-inverse; this is required as the inverse of $\matA_{0}-\Omega_{0}^2\matB_{0}$ is not unique; we check that this non-uniqueness has no impact upon the final solution, which is ensured by \eqref{eq:provided}. Applying the solvability condition at the next order, we arrive at

\begin{align}
\begin{split}
\Omega_{2}^2 = \frac{\kappa\mathbf{W}_{0}^{\dagger}\left(\matA_{1}-\Omega_{0}^2\matB_{1}\right)\mathbf{W}_{1} + \frac{\kappa^2}{2}\mathbf{W}_{0}^{\dagger}\left(\matA_{2}-\Omega_{0}^2\matB_{2}\right)\mathbf{W}_{0}}{\mathbf{W}_{0}^\dagger\matB_{0}\mathbf{W}_{0}}.
\end{split}
\end{align}

Requiring the Bloch conditions are satisfied on the macroscale, we set $f(X) = \exp(i\kappa X)$ for the periodic case on the microscale (as it is at $\Gamma$), and using $T = \Omega_2^2/\kappa^2 $ we obtain 

\begin{align}
T = \frac{-\mathbf{W}_{0}^{\dagger}\matC_{1}\matC_{0}^{+}\matC_{1}\mathbf{W}_{0} + \frac{1}{2}\mathbf{W}_{0}^{\dagger}\matC_{2}\mathbf{W}_{0}}{\mathbf{W}_{0}^\dagger\matB_{0}\mathbf{W}_{0}} 
\end{align}
with
\begin{align}
\begin{split}
    \matC_{i} = \left(\matA_{i} - \Omega_{0}^2\matB_{i}\right).
\end{split}
\end{align}

\clearpage

\bibliographystyle{elsarticle-num} 

\end{document}